\documentclass[twocolumn,showpacs,amssymb,aps,prc,nofootinbib,floatfix,superscriptaddress]{revtex4-1}
\usepackage{amsmath,graphicx,color,ulem}
\usepackage{hyperref}
\newcommand{\trento}{T$\mathrel{\protect\raisebox{-2.1pt}{R}}$ENTo}

\begin{document}
\title{Probing collectivity in heavy-ion collisions with fluctuations of the $p_T$ spectrum}

\author{Tribhuban Parida}
\affiliation{Department of Physical Sciences,
Indian Institute of Science Education and Research Berhampur, Transit Campus (Govt ITI), Berhampur-760010, Odisha, India}
\author{Rupam Samanta}
\affiliation{AGH University of Krakow, Faculty of Physics and
Applied Computer Science, aleja Mickiewicza 30, 30-059 Cracow, Poland}
\author{Jean-Yves Ollitrault}
\affiliation{Universit\'e Paris Saclay, CNRS, CEA, Institut de physique th\'eorique, 91191 Gif-sur-Yvette, France}

\begin{abstract} 
Event-by-event fluctuations in the initial stages of ultrarelativistic nucleus-nucleus collisions depend little on rapidity. 
The hydrodynamic expansion which occurs in later stages then gives rise to correlations among outgoing particles which depend weakly on their relative rapidity. 
Azimuthal correlations, through which anisotropic flow  ($v_n(p_T)$) is defined, have been the most studied.
Here we study a new observable introduced in 2020 by Schenke, Shen  and Teaney and dubbed $v_0(p_T)$, which quantifies the relative change in the $p_T$ spectrum induced by a fluctuation. 
We describe how it can be measured. 
Using hydrodynamic simulations, we make quantitative predictions for $v_0(p_T)$ of charged and identified hadrons. 
We then discuss how $v_0(p_T)$ relates to two phenomena which have been measured: 
The increase of the mean transverse momentum in ultracentral collisions, and the event-by-event fluctuations of the transverse momentum per particle $[ p_T]$.
We show that $v_0(p_T)$ determines the dependence of these quantities on the $p_T$ cuts implemented in the analysis.
We quantitatively explain the rise of $\sigma_{p_T}$ observed by ATLAS as the upper $p_T$-cut is increased from $2$ to $5$~GeV/$c$. 
  \end{abstract}

\maketitle

\section{Introduction}
Evidence for the formation of a thermalized fluid in ultrarelativistic nucleus-nucleus collisions  relies on the observation of correlations between outgoing hadrons at large relative rapidity (long-range correlations)~\cite{Alver:2010gr,Ollitrault:2023wjk}. 
Their interpretation is the following:
In a single collision event, the initial density profile depends little on the longitudinal coordinate (space-time rapidity)~\cite{Bjorken:1982qr,Dumitru:2008wn} but has in general an irregular shape in the transverse plane, generated by quantum fluctuations in the wavefunctions of incoming nuclei~\cite{PHOBOS:2006dbo}. 
The hydrodynamic expansion carries over the initial pattern to the momentum distribution of outgoing hadrons through pressure gradients~\cite{Ollitrault:1992bk}. 
The most spectacular manifestation of collectivity is the  large elliptic flow in non-central collisions~\cite{STAR:2000ekf}, generated by the almond shape of the overlap between the nuclei. 
More generally, anisotropic flow~\cite{ALICE:2011ab} is generated by the azimuthal anisotropy of the initial density profile, a phenomenon referred to as   ``shape-flow transmutation''~\cite{Bozek:2014cva}. 

What is less often realized is that there are also long-range correlations which do not involve the azimuthal angle. 
The transverse momenta of outgoing particles are also correlated~\cite{PHENIX:2003ccl,STAR:2005xhg,ALICE:2014gvd,ATLAS:2022dov}.
The physical interpretation is that the transverse momentum per particle, $[p_T]$, fluctuates event to event, and that these fluctuations are not only statistical. 
The standard deviation of dynamical fluctuations, denoted by $\sigma_{p_T}$, is of the order of $1\%$ in central Pb+Pb collisions at the LHC~\cite{ATLAS:2022dov}.  
In hydrodynamics, when comparing events with the same final multiplicity, those with larger $[p_T]$ correspond to more compact initial density profiles~\cite{Broniowski:2009fm}, with smaller transverse radius. 
The larger $[p_T]$ is generated by larger pressure gradients. 
Close scrutiny of the multiplicity dependence of $\sigma_{p_T}$ in ultracentral collisions~\cite{ATLAS:2022dov} confirms that this mechanism is at work at the LHC~\cite{Samanta:2023amp}. 
Longitudinal symmetry of the initial density profile implies that the fluctuations of $[p_T]$ induce transverse momentum correlations which are  independent of the relative pseudorapidity $\Delta\eta$, in the same way as azimuthal correlations~\cite{CMS:2012xss}. 
However, we are not aware of any experimental measurement of their variation with $\Delta\eta$. 

In this paper, we study the event-by-event fluctuations of $[p_T]$ differentially in $p_T$. 
That is, we evaluate the fluctuation of the $p_T$ spectrum associated with a fluctuation of $[p_T]$. 
A new observable has been coined to measure this effect and dubbed $v_0(p_T)$ \cite{Schenke:2020uqq}. 
It has not yet been analyzed in heavy-ion experiments. 
We will show that $v_0(p_T)$ corresponds to the relative change in the spectrum generated by a small increase in the initial temperature. 
For a given multiplicity, more compact events have higher initial temperature, so that this interpretation is compatible with the traditional picture of size-flow transmutation~\cite{Bozek:2017elk}. 
This interpretation can be more intuitive: 
In hydrodynamics, a system with a larger initial temperature will reach a higher fluid velocity. 
The fluid velocity at freeze-out determines the spectra of outgoing hadrons~\cite{Cooper:1974mv,Schnedermann:1993ws,Guillen:2020nul}. 
Thus, $v_0(p_T)$ is the observable which quantifies the change in the spectrum induced by a change in the fluid velocity. 

We recall the definition of $v_0(p_T)$ in Sec.~\ref{s:definitions} and derive the sum rules that relate it to $\sigma_{p_T}$.  
We introduce the scaled observable $v_0(p_T)/v_0$, which will be shown to be largely independent of system size and centrality at a given collision energy, in the same way as the scaled anisotropic flow $v_n(p_T)/v_n$~\cite{ATLAS:2018ezv} and scaled $p_T$ spectra~\cite{Muncinelli:2024izj}. 
We then carry out a systematic study of $v_0(p_T)/v_0$ in hydrodynamics. 
We study its sensitivity to initial state fluctuations by running two independent sets of hydrodynamic simulations, which are presented in Sec.~\ref{s:hydro}. 
The first set is a standard simulation with fluctuating initial conditions (IC)~\cite{Holopainen:2010gz,Schenke:2010rr}, mimicking an actual experiment. 
The second set uses smooth initial conditions, where we study the effect of a small change in the initial temperature by slightly increasing the initial entropy density by a global factor close to unity. 
The goal is to test whether initial temperature fluctuations are the main factor driving $v_0(p_T)$. 
We study, for these two sets of calculations, the sensitivity of $v_0(p_T)/v_0$ to the transport coefficients (shear and bulk viscosities) of the quark gluon plasma. 
Our results are presented  in Sec.~\ref{s:results}, for unidentified charged particles and identified hadrons. 
In Sec.~\ref{s:acceptance}, we study quantitatively the effect of upper and lower cuts in the $p_T$ acceptance, which differ depending on the detector, on the transverse momentum per particle. 
We derive the correction to $\sigma_{p_T}$ stemming from these cuts, which we express as a function of $v_0(p_T)$.
Several applications are discussed in Sec.~\ref{s:applications}.
We show that our hydrodynamic calculation reproduces quantitatively the increase by a factor $\sim 2$ of $\sigma_{p_T}$ observed by ATLAS as the upper cut is raised from $2$ to $5$~GeV/$c$~\cite{ATLAS:2022dov} (Sec.~\ref{s:ATLAS}).
We also explain PHENIX data at lower energy~\cite{PHENIX:2003ccl}, which are however less precise (Sec.~\ref{s:PHENIX}). 
Finally, we argue that the same dependence on $p_T$ cuts applies to the increase of $\langle p_T\rangle$ in ultracentral collisions~\cite{Gardim:2019brr,Nijs:2023bzv,CMS:2024sgx} (Sec.~\ref{s:cs}).
An experimental procedure to analyze $v_0(p_T)$ is suggested in Appendix~\ref{s:analysis}.

\section{Definitions}
\label{s:definitions}

We follow Schenke, Shen and Teaney in their definition of $v_0(p_T)$~\cite{Schenke:2020uqq}.
For the analysis, one typically sorts the particles in bins of $p_T$, covering the acceptance of the detector. 
For a given event, we denote by $N(p_T)$ the number of particles in a $p_T$ bin, i.e., the $p_T$ spectrum.
The total number of particles, $N$, and the transverse momentum per particle,  $[p_T]$, are given by 
\begin{eqnarray}
  \label{sumrule1}
  N & = & \int_{p_T} N(p_T)\cr 
  [p_T] & = & \frac{1}{N}\int_{p_T} p_TN(p_T). 
\end{eqnarray}
where $\int_{p_T}$ denotes the sum over all bins.

We then consider an ensemble of events, corresponding typically to a centrality class.
We denote by $N_0(p_T)$ the average value of $N(p_T)$ over all events, 
by $N_0$ the average multiplicity, and by $\langle p_T\rangle$ the value of $[p_T]$ corresponding to this average spectrum.
For the average spectrum, equations similar to Eqs.~(\ref{sumrule1}) give: 
\begin{eqnarray}
  \label{sumruleav}
  N_0 & = & \int_{p_T} N_0(p_T)\cr
  \langle p_T\rangle & = & \frac{1}{N_0}\int_{p_T} p_TN_0(p_T). 
\end{eqnarray}
We then carry out a fluctuation decomposition of $N(p_T)$, $N$ and $[p_T]$:
\begin{eqnarray}
  \label{deffluct}
  N(p_T)&=& N_0(p_T)+\delta N(p_T)\cr 
  N&=& N_0+\delta N\cr 
 [p_T]&=& \langle p_T\rangle+\delta p_T. 
\end{eqnarray}
These equations define $\delta N(p_T)$, $\delta N$ and $\delta p_T$. 
$\delta N$ and $\delta p_T$ can also be expressed as integrals of $\delta N(p_T)$: 
\begin{eqnarray}
  \label{decomposition}
  \delta N&=&\int_{p_T} \delta N(p_T)\cr
  \delta p_T&=&\frac{1}{N_0}\int_{p_T}\left( p_T-\langle p_T\rangle\right) \delta N(p_T),
\end{eqnarray}
where the second equation is obtained by linearizing in the fluctuations. 
This decomposition will be used below. 

The quantities of interest in this paper, $\sigma_{p_T}$ and $v_0(p_T)$, are obtained by first selecting events with the same multiplicity, that is, $\delta N=0$. 
They are defined by: 
\begin{equation}
  \label{defsigma}
\sigma_{p_T}^2\equiv \langle (\delta p_T)^2\rangle, 
\end{equation}
and 
\begin{equation}
  \label{defv0pt}
v_0(p_T)\equiv\frac{\langle \delta N(p_T)\delta p_T\rangle}{N_0(p_T) \sigma_{p_T}}, 
\end{equation}
where angular brackets denote an average over events. 
Thus $v_0(p_T)$ represents the correlation between a fluctuation in the spectrum, and a fluctuation in the transverse momentum per particle. 
Using Eq.~(\ref{decomposition}), one obtains the following sum rules: 
\begin{eqnarray}
  \label{sumrules3}
\int_{p_T}  v_0(p_T) N_0(p_T)&=&0\cr
\int_{p_T} (p_T-\langle p_T\rangle) v_0(p_T) N_0(p_T)&=&\sigma_{p_T} N_0.  
\end{eqnarray}
The first equation implies that $v_0(p_T)$  changes sign as a function of $p_T$,
like directed flow, $v_1(p_T)$~\cite{Luzum:2010fb},\footnote{The change of sign of $v_1(p_T)$ is implied by the condition that the total net transverse momentum vanishes.} and unlike elliptic flow and higher harmonics, $v_n(p_T)$ with $n\ge 2$, which are typically positive for all $p_T$. 

Following Ref.~\cite{Schenke:2020uqq} again, we define the ``integrated'' $v_0$ by
\begin{equation}
\label{defv0}
  v_0\equiv \frac{\sigma_{p_T}}{\langle p_T\rangle}. 
\end{equation}
Throughout this paper, we study the ``scaled  $v_0(p_T)$'' defined by:
\begin{equation}
  \label{defscaledv0}
  \frac{v_0(p_T)}{v_0}\equiv
\frac{\langle p_T\rangle}{N_0(p_T)}  \frac{\langle \delta N(p_T)\delta p_T\rangle}{\sigma_{p_T}^2}. 
\end{equation}
The interest of this scaled variable is that it is independent of the size of the fluctuations, as defined by $\sigma_{p_T}$. 
Thus the dependence on models of initial conditions is largely suppressed, as will be shown in Sec.~\ref{s:results}. 

The sum rules (\ref{sumrules3}) can be rewritten, using Eq.~(\ref{defv0}): 
\begin{eqnarray}
  \label{sumrules4}
\int_{p_T}  \frac{v_0(p_T)}{v_0} N_0(p_T)&=&0\cr
\int_{p_T} (p_T-\langle p_T\rangle) \frac{v_0(p_T)}{v_0} N_0(p_T)&=&\langle p_T\rangle N_0.  
\end{eqnarray}

\section{Hydrodynamic setup}
\label{s:hydro}

Schenke, Shen and Teaney have already carried out state-of-the-art hydrodynamic simulations and made quantitative predictions for $v_0(p_T)$~\cite{Schenke:2020uqq} for specific centralities. 
Our goal is to carry out a more systematic study, albeit within a less sophisticated setup, as will be detailed below. 
Like these authors, we assume longitudinal boost invariance~\cite{Bjorken:1982qr}. 
Therefore, any fluctuation in the spectrum is independent of rapidity, and induces a long-range correlation which is independent of the rapidity gap. 
We simulate Pb+Pb collisions at $\sqrt{s_{\rm NN}}=5.02$~TeV, the energy of the second run of the LHC~\cite{ALICE:2015juo}. 

We carry out two sets of simulations. 
The first set uses fluctuating initial conditions. 
We generate 100  initial density profiles which are then evolved through hydrodynamics. 
We then treat these events using the exact same procedure as if they were experimental events, described in Sec.~\ref{s:definitions}. 
The only notable difference with an actual experiment is that in hydrodynamics, one calculates the probability distribution of momentum at freeze-out, which is continuous and need not be sampled (unless one wants to model subsequent rescatterings, a point to which we come back). 
Therefore, there are no fluctuations coming from the finite multiplicity, and a much smaller number of events is needed than in an actual experiment in order to achieve reasonable statistical accuracy~\cite{Gardim:2011qn}. 

The second set of simulations uses a unique, smooth initial entropy density profile, obtained by averaging over $10^4$ fluctuating initial conditions~\cite{Song:2010mg}.\footnote{Before averaging, we rotate each event by its respective second-order participant plane angle to align the second-order participant plane of each event and set the second-order participant plane to zero in each event~\cite{Shen:2020jwv}.}
We run two calculations with this initial density profile, which differ only by a global factor $1.05$. 
Thus we obtain two ``events'', whose $[p_T]$ differ by $\delta p_T$. 
Equation~(\ref{defsigma}) then gives $\sigma_{p_T}=|\delta p_T|$, and Eq.~(\ref{defscaledv0}) can be rewritten as:
\begin{equation}
\label{defv0smooth}
\frac{v_0(p_T)}{v_0}\equiv \frac{\delta\ln N(p_T)}{\delta\ln [p_T]},
\end{equation}
where $\delta X$ denotes the (small) difference of observable $X$ between the two events. 

By default, we simulate head-on collisions (zero impact parameter), unless otherwise specified. 
But the scaled quantity $v_0(p_T)/v_0$ turns out to be  constant from central to mid-central collisions, as will be shown explicitly below.  
The underlying reason is that scaled $p_T$ spectra are independent of centrality~\cite{Muncinelli:2024izj} and that $\langle p_T\rangle$ itself varies by less than 1\% in the $0-30\%$ centrality window~\cite{ALICE:2018hza}. 
For more peripheral collisions, $v_0(p_T)/v_0$ should still be approximately same (not much difference from central collisions) when expressed as a function of $p_T/\langle p_T\rangle$. 

We start the hydrodynamic evolution at proper time $\tau_0=0.4$~fm$/c$, and transverse flow at earlier times~\cite{Vredevoogd:2008id,Kurkela:2018wud} is neglected (note that it is taken into account in the IP-Glasma initial conditions of Ref.~\cite{Schenke:2020uqq}). 
The initial entropy density is given by the \trento{} model of initial conditions~\cite{Moreland:2014oya}. 
Within this model, we implement the usual prescription with $p=0$, which is favored by model to data comparisons~\cite{Bernhard:2019bmu,JETSCAPE:2020mzn,Nijs:2020roc}, and corresponds to an density profile proportional to $\sqrt{T_AT_B}$, where $T_A$ and $T_B$ are the thickness functions of colliding nuclei. 
The fluctuation parameter is $k=2.5$. 
We normalize the density profile so as to reproduce the final multiplicity measured by ALICE in the $0-5$\% centrality window~\cite{ALICE:2015juo}. 

We solve hydrodynamic equations  using the publicly available MUSIC code~\cite{Schenke:2010nt,Schenke:2010rr,Paquet:2015lta}, with the Huovinen-Petreczky lattice-QCD equation of state with partial chemical equilibrium~\cite{Huovinen:2009yb}. 
For each set of simulations, we run ideal and viscous hydrodynamic calculations. 
Note that in the case of fluctuating initial conditions, calculations with different viscosities are run with the same initial profiles (upon adjusting the normalization to achieve a constant final multiplicity) so as to isolate the systematic effect of varying the viscosity~\cite{Noronha-Hostler:2015uye}. 
In order to assess the respective effects of shear and bulk viscosity on observables, we implement them in separate calculations. 
For shear viscosity, we assume a constant shear viscosity over entropy ratio $\eta/s=0.08$ or $\eta/s=0.16$~\cite{Romatschke:2007mq}. 
For bulk viscosity, we use the temperature-dependent  parametrization of $\zeta/s(T)$ by the McGill group~\cite{Ryu:2015vwa,Paquet:2015lta,Ryu:2017qzn}. 
We perform particlization following the Cooper-Frye prescription~\cite{Cooper:1974mv} on a freezeout hypersurface at $T=130$~MeV (corresponding to an energy density $0.18$~GeV/fm$^3$). 
A more sophisticated procedure~\cite{Schenke:2020mbo} is to implement freeze-out at a higher temperature, and couple the fluid evolution to a transport code in order to model hadronic interactions. 
The advantage of our modeling is that we need not sample particles with a Monte Carlo algorithm at freeze-out. 
Our calculations are done directly using the continuous momentum distribution defined by the Cooper-Frye prescription, so that there is no statistical error associated with hadronization. 
The price to pay is that the description of identified-particle spectra is less precise, as will be detailed in Sec.~\ref{s:results}. 
Note, finally, that there is some arbitrariness in the choice of the viscous correction to the momentum distribution at freeze-out~\cite{Teaney:2003kp,Dusling:2011fd,McNelis:2021acu}. 
We choose the Chapman-Enskog prescription~\cite{Paquet:2015lta,Ryu:2017qzn}.

\section{Results}
\label{s:results}

\begin{figure}[th!]
    \includegraphics[width=\linewidth]{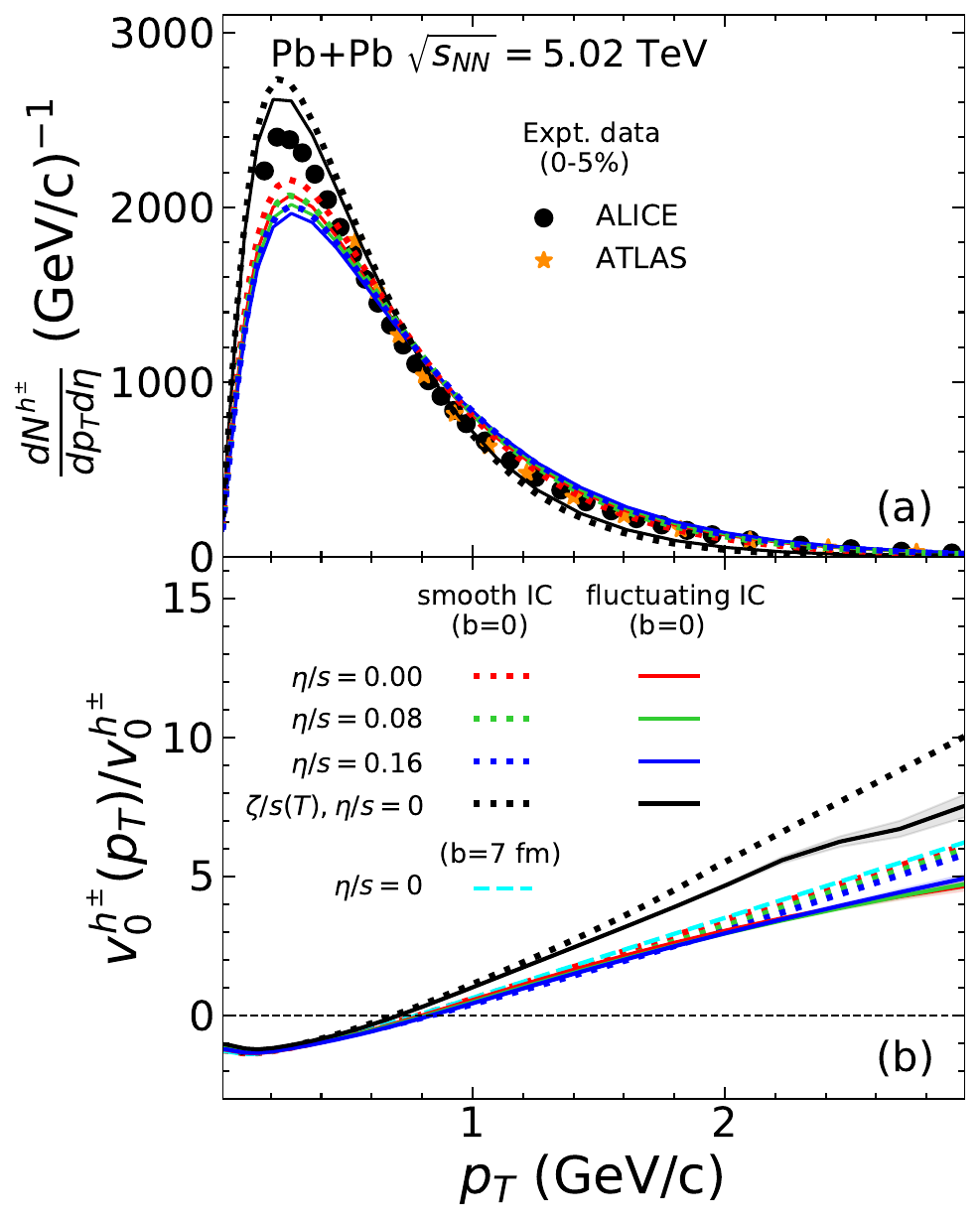}
	\caption{
	Hydrodynamic predictions for 
	(a) Symbols: spectra of unidentified charged particles for Pb+Pb collisions at $\sqrt{s_{\rm NN}}=5.02$~TeV in the 0-5\% centrality window measured by ALICE~\cite{ALICE:2018vuu} ($|\eta|<0.8$) and ATLAS ~\cite{ATLAS:2022kqu} ($|\eta|<2.5$). Lines: hydrodynamic calculations with smooth (dotted) or fluctuating (full) initial conditions.  
	(b) Hydrodynamic predictions for $v_0(p_T)/v_0$. } 
        \label{fig:v0ptbyv0charged}
\end{figure}
Fig.~\ref{fig:v0ptbyv0charged} (a) displays spectra of unindentified charged particles. 
One first notes that for a given choice of transport coefficients, results are very similar for smooth initial conditions and fluctuating initial conditions. 
Fluctuations increase somewhat $dN/dp_T$ at  high-$p_T$~\cite{Andrade:2008xh}, and decrease it at low $p_T$. 
Shear viscosity increases the mean $p_T$~\cite{Romatschke:2007mq}: 
Like fluctuations, it makes the spectrum harder.
The effect of bulk viscosity is larger in magnitude than that of shear viscosity, and goes in the opposite direction: 
It softens the spectrum~\cite{Bozek:2009dw,Noronha-Hostler:2013gga,Ryu:2015vwa}. 
We also display experimental data from ALICE~\cite{ALICE:2018vuu} and ATLAS~\cite{ATLAS:2022kqu} to show that our hydrodynamic calculation is in the ballpark. 
Note that our hydrodynamic calculation is for  the spectrum per unit rapidity $y$, while experimental data are per unit {\it pseudorapidity\/} $\eta$. 
We neglect the small Jacobian correction between these two quantities, as it would be different for ALICE and ATLAS. 

Fig.~\ref{fig:v0ptbyv0charged} (b) displays our hydrodynamic predictions for $v_0(p_T)/v_0$. 
$v_0(p_T)/v_0$ changes sign as a function of $p_T$, as expected from Eq.~(\ref{sumrules4}). 
It is positive for high $p_T$, which is a direct consequence of the definition (\ref{defv0pt}): 
An upward fluctuation in the transverse momentum per particle, $\delta p_T>0$, goes along with an increase of the number of high-$p_T$ particles, that is, a positive  $\delta N(p_T)$. 
The $p_T$ dependence is similar to that of directed flow, $v_1(p_T)$~\cite{Retinskaya:2012ky}, with the notable difference that $v_1(p_T)$ is proportional to $p_T$ in the  limit $p_T\to 0$, while $v_0(p_T)$ converges to a finite negative value.

First, one notes that the two sets of calculations, with initial fluctuations and with smooth initial conditions, return very similar results. 
With smooth initial conditions, $v_0(p_T)/v_0$ is solely driven by the fluctuation in the initial temperature. 
This in turn implies that  temperature fluctuations are the key phenomenon driving $v_0(p_T)$, also with fluctuating initial conditions. 

An important point is that by studying the scaled flow $v_0(p_T)/v_0$ instead of just $v_0(p_T)$, dependence on model parameters is greatly reduced. 
The fact that the two sets of calculations give almost identical results implies that $v_0(p_T)/v_0$ is largely independent of the modeling of the fluctuating initial density profile. 
By contrast, $v_0$ alone (hence $v_0(p_T)$) strongly depends on initial fluctuations. 
According to Eq.~(\ref{defv0}), it is proportional to $\sigma_{p_T}$, which is a key observable in constraining the initial state from experimental data~\cite{Bozek:2017elk,Bernhard:2019bmu,JETSCAPE:2020mzn,Nijs:2020roc}. 

To an excellent approximation, $v_0(p_T)/v_0$ is also independent of centrality. 
In order to illustrate this, we also display the result of the smooth ideal hydrodynamic calculation for an impact parameter $b=7$~fm. 
The multiplicity is typically smaller by a factor two than for $b=0$, but $v_0(p_T)/v_0$ is identical. 
There is a tiny dependence on centrality in viscous hydrodynamics (not shown), because the relative change of $v_0(p_T)/v_0$ due to viscosity is inversely proportional to the transverse radius, according to Reynolds number scaling~\cite{Gardim:2020mmy}. 
But this is a negligible effect for the range of centralities where hydrodynamics is expected to apply~\cite{Muncinelli:2024izj}.

The dependence on transport coefficients is also reduced by calculating $v_0(p_T)/v_0$. 
The residual dependence on initial-state fluctuations and transport coefficients is, at least qualitatively, the same as for anisotropic flow $v_n(p_T)$. 
Fluctuations and shear viscosity reduce  $v_0(p_T)/v_0$ at high $p_T$ in the same way as they reduce $v_2$ at high $p_T$~\cite{Andrade:2008xh,Teaney:2003kp,Romatschke:2007mq},  while bulk viscosity increases it~\cite{Noronha-Hostler:2013gga}. 
Note that the dependence of $v_0(p_T)/v_0$ on transport coefficients results, at least partly, from  that of $\langle p_T\rangle$, since $v_0(p_T)/v_0$ is related to the spectrum through the sum rules (\ref{sumrules4}).  

%One sees that $v_0(p_T)/v_0$  is much more sensitive to bulk viscosity than to shear viscosity. 
%so that this observable could help constrain these two transport coefficients separately. 

\begin{figure}[th!]
	\includegraphics[width=\linewidth]{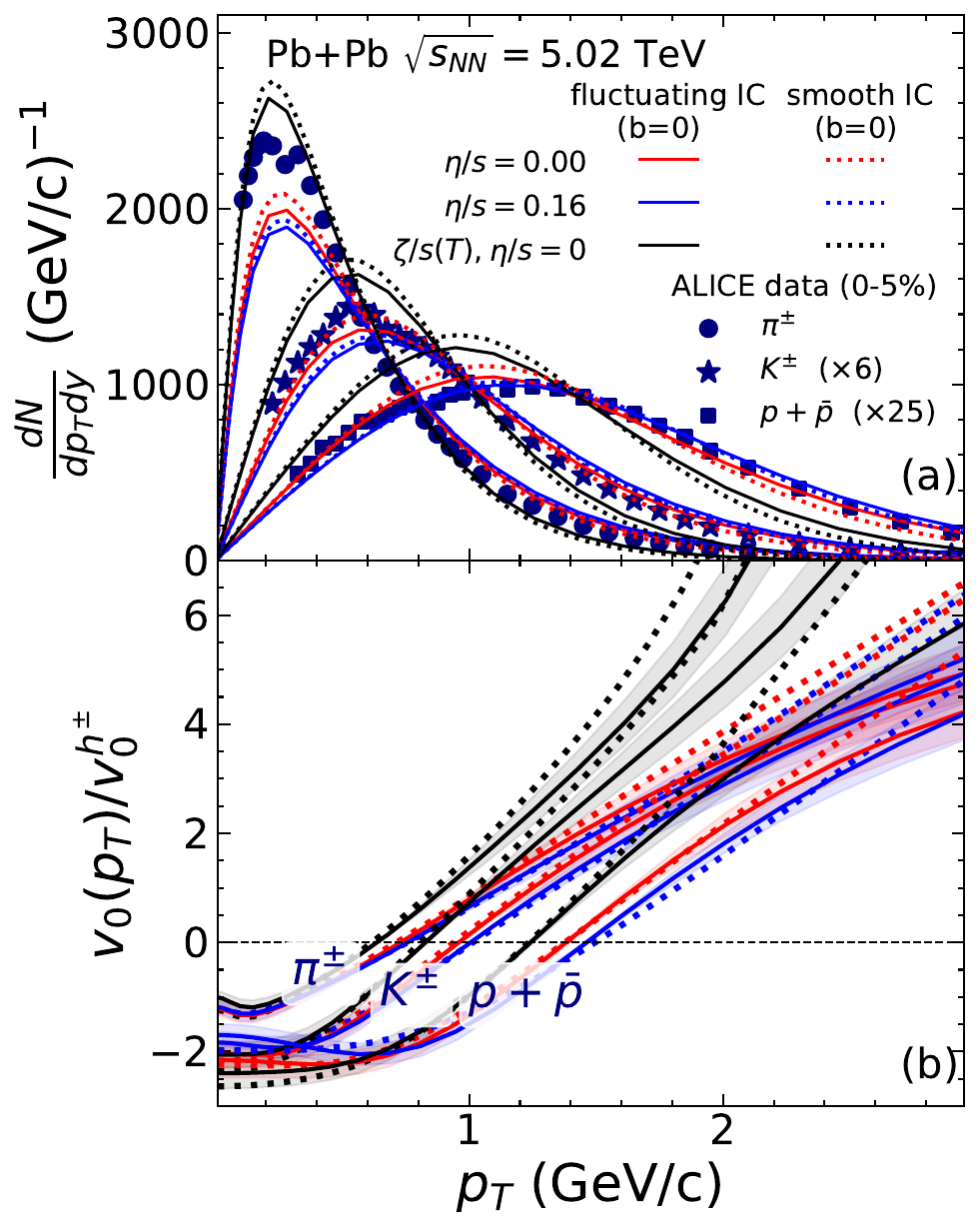} 
	\caption{ 
	(a) Spectra of identified pions, kaons and protons for Pb+Pb collisions at $\sqrt{s_{\rm NN}}=5.02$~TeV in the 0-5\% centrality window.
	Symbols: ALICE data~\cite{ALICE:2019hno}
	Lines: hydrodynamic calculations with smooth (dotted) or fluctuating (full) initial conditions.  
	(b) Hydrodynamic predictions for $v_0(p_T)/v_0$ of identified pions, kaons and protons. 
 Note that the denominator $v_0$ is that of all charged particles, not of identified particles (see text). Shaded bands correspond to the statistical error
in the calculation with fluctuating initial conditions. 
 }               
	\label{fig:v0ptbyv0idpart}
        \end{figure}
Fig.~\ref{fig:v0ptbyv0idpart} (a) displays spectra of   identified pions, kaons and protons. 
The effect of fluctuations and transport coefficients is qualitatively the same as for charged particle spectra in Fig.~\ref{fig:v0ptbyv0charged}. 
One sees that hydrodynamics generally underestimates the transverse momentum of protons compared to that of pions and kaons. 
It is a well-known feature which can be improved~\cite{Ryu:2017qzn} by taking into account hadronic rescatterings after the end of the hydrodynamic evolution~\cite{Petersen:2008dd,SMASH:2016zqf}. 

Fig.~\ref{fig:v0ptbyv0idpart} (b) presents our predictions for $v_0(p_T)/v_0$ of identified particles. 
We choose to scale by the ``integrated'' $v_0^{h^\pm}$ for all charged particles, which is easier to measure than that of identified particles. 
With this definition, the denominator is the same for pions, kaons and protons, and only the numerator $v_0(p_T)$ depends on the particle species. 
Therefore, the ordering between the curves is the same as if we had plotted $v_0(p_T)$. 
$v_0(p_T)$ displays a clear mass ordering, in the sense that  at a given value of $p_T$, it decreases with increasing particle mass~\cite{Schenke:2020uqq}. 
Hydrodynamics predicts a similar mass ordering for elliptic flow, $v_2(p_T)$~\cite{Huovinen:2001cy}, which is observed at RHIC~\cite{STAR:2001ksn,PHENIX:2003qra} and LHC~\cite{ALICE:2014wao}. 
The same phenomenon is observed for higher harmonics of anisotropic flow, $v_n(p_T)$ with $n\ge 3$~\cite{PHENIX:2014uik,ALICE:2016cti}. 
The mass ordering can be understood simply as follows in ideal hydrodynamics~\cite{Borghini:2005kd,Ollitrault:2007du}: 
Assuming that the particle momentum is parallel to the fluid velocity at freeze-out, the momentum distribution in a fluid cell is a boosted Boltzmann distribution $\propto \exp((p_Tu-m_tu_0)/T)$, where $m_t\equiv \sqrt{p_T^2+m^2}$ is the transverse mass and  $u_0=\sqrt{1+u^2}$. 
Now, $v_0(p_T)$ corresponds to the relative variation of the spectrum upon a small fluctuation change in the 4-velocity $u$. 
Using $u_0\delta u_0=u\delta u$ and applying Eq.~(\ref{defv0smooth}), one immediately obtains
\begin{equation}
\label{massordering}
\frac{v_0(p_T)}{v_0}\propto (p_T-m_t v), 
\end{equation}
where $v\equiv u/u_0$ is the fluid velocity. 
Thus, for a given $p_T$,  $v_0(p_T)$ decreases as the particle mass increases.

\section{Acceptance correction}
\label{s:acceptance}

Analyses are usually done using an interval of $p_T$ which does not extend from zero to infinity, but covers a finite interval $(p_T)_{\min}<p_T<(p_T)_{\max}$.
We now study the effect of such a cut on the momentum per particle $[p_T]$, and show how $v_0(p_T)/v_0$ can be used in order to capture it. 

We need to make assumptions as to the fluctuation of the spectrum $\delta N(p_T)$, from which $[p_T]$ is defined. 
We assume that the relative fluctuation can be decomposed as a global increase proportional to $\delta N$, and a term proportional to $\delta p_T$:
\begin{equation}
\label{deltanpt}
\frac{\delta N(p_T)}{N_0(p_T)}=\frac{\delta N}{N_0}+v_0(p_T)\frac{\delta p_T}{\sigma_{p_T}}. 
\end{equation}
Setting $\delta N=0$ in this equation, multiplying with $\delta p_T$, averaging over events and using Eq.~(\ref{defsigma}), one sees that it is compatible with the definition of $v_0(p_T)$ in Eq.~(\ref{defv0pt}).

The decomposition (\ref{deltanpt}) amounts to assuming that the fluctuation of the spectrum  can be described as a superposition of two principal components~\cite{Bhalerao:2014mua}.
The analysis by CMS has shown that this is a very good approximation~\cite{CMS:2017mzx}. 
Then, these two components can be interpreted as coming from the event-by-event fluctuations of $N$ and $[p_T]$~\cite{Gardim:2020fxx}, which corresponds to the decomposition (\ref{deltanpt}). 

Note that the first term in the right-hand side of Eq.~(\ref{deltanpt}) is independent of $p_T$. 
It corresponds to a global rescaling of the spectrum, without any change in the slope (i.e., $\delta p_T=0$). 
The second term corresponds to a fluctuation of $[p_T]$ at fixed multiplicity, which goes along with a non-trivial modification of the $p_T$ spectrum. 
The observable $v_0(p_T)$ quantifies how this fluctuation depends on $p_T$. 

Using Eq.~(\ref{defv0}), one can rewrite Eq.~(\ref{deltanpt}) in terms of $v_0(p_T)/v_0$, which is the quantity we are calculating: 
\begin{equation}
\label{deltanpt2}
\frac{\delta N(p_T)}{N_0(p_T)}=\frac{\delta N}{N_0}+ \frac{v_0(p_T)}{v_0} \frac{\delta p_T}{\langle p_T\rangle}.
\end{equation} 

Let us now study the effects of a finite acceptance, where integrals over $p_T$ are restricted to an interval. 
We denote generically this condition by  $p_T\in {\cal A}$. 
The acceptance has an effect on integrated quantities, not on the spectrum $N(p_T)$ itself, nor on its fluctuation $\delta N(p_T)$. 

On the other hand, all equations involving integrals must be rewritten: 
\begin{eqnarray}
  \label{sumrulesA}
N_{\cal A}&\equiv &\int_{p_T\in {\cal A}} N(p_T)\cr
[p_T]_{\cal A}&\equiv&\frac{1}{N_{\cal A}}\int_{p_T\in {\cal A}} p_T N(p_T),
\end{eqnarray}
Let us denote by $N_{0{\cal A}}$ and $\langle p_T\rangle_{\cal A}$ the values of $N_0$ and $\langle p_T\rangle$ within the acceptance:
\begin{eqnarray}
  \label{sumrulesavA}
N_{0{\cal A}}&\equiv &\int_{p_T\in {\cal A}} N_0(p_T)\cr
\langle p_T\rangle_{\cal A}&\equiv&\frac{1}{N_{0{\cal A}}}\int_{p_T\in {\cal A}} p_T N_0(p_T), 
\end{eqnarray}
and by $\delta N_{\cal A}$ and $\delta p_{T,A}$ the event-by-event fluctuations of $N_{\cal A}$ and $[p_T]_{\cal A}$:  
\begin{eqnarray}
  \label{deffluctA}
N_{\cal A}&=&N_{0{\cal A}}+\delta N_{\cal A}\cr
[p_T]_{\cal A}&=& \langle p_T\rangle_{\cal A}+\delta p_{T,A}. 
\end{eqnarray}
The algebra is identical to that of Sec.~\ref{s:definitions}, and Eq.~(\ref{decomposition}) thus becomes: 
\begin{eqnarray}
  \label{decompositionA}
\delta N_{\cal A}&=&\int_{p_T\in {\cal A}} \delta N(p_T)\cr
\delta p_{T,\cal A}&=&\frac{1}{N_{0\cal A}}\int_{p_T\in {\cal A}} (p_T -\langle p_T\rangle_{\cal A})\delta N(p_T).
\end{eqnarray}

Inserting Eq.~(\ref{deltanpt2}) into the second equation, one find that the first term does not contribute. 
Evaluating the second term, one obtains
\begin{equation}
\label{ptacc}
\frac{\delta p_{T,\cal A}}{\langle p_T\rangle_{\cal A}}= C_{\cal A}\frac{\delta p_T}{\langle p_T\rangle},
\end{equation}
where $C_{\cal A}$ is a dimensionless correction factor due to acceptance, whose expression is: 
\begin{eqnarray}
  \label{defCA}
C_{\cal A}&\equiv &\frac{1}{N_{0{\cal A}}\langle p_T\rangle_{\cal A}}\int_{p_T\in {\cal A}} (p_T-\langle p_T\rangle_{\cal A})\frac{v_0(p_T)}{v_0}N_0(p_T).
\end{eqnarray}
If all the $p_T$ range is covered by the acceptance, then by applying Eqs.~(\ref{sumrules4}), one immediately obtains $C_{\cal A}=1$, as one should.

First, it is important to note that $v_0(p_T)$ is insensitive to the transverse momentum cuts, much as differential anisotropic flow $v_n(p_T)$.
The reason is that $\sigma_{p_T}$, defined by Eq.~(\ref{defsigma}), is proportional to $\delta p_T$, so that Eq.~(\ref{ptacc}) gives: 
\begin{equation}
\label{ptacc2}
\frac{\sigma_{p_T,\cal A}}{\langle p_T\rangle_{\cal A}}= C_{\cal A}\frac{\sigma_{p_T}}{\langle p_T\rangle}. 
\end{equation}
As a consequence, the factor $C_{\cal A}$ cancels between the numerator and the denominator of Eq.~(\ref{defv0pt}).
On the other hand, the ``integrated'' $v_0$ depends on the acceptance cuts.
Inspection of Eqs.~(\ref{defv0}) and (\ref{ptacc2}) gives immediately: 
\begin{equation}
\label{ptacc3}
v_{0{\cal A}} = C_{\cal A}v_0.
\end{equation}
That is, the integrated $v_0$ is also multiplied by $C_{\cal A}$. 
Thus the acceptance correction increases the ratio $v_0(p_T)/v_0$, through the denominator.

\begin{figure}[th!]
	\includegraphics[width=\linewidth]{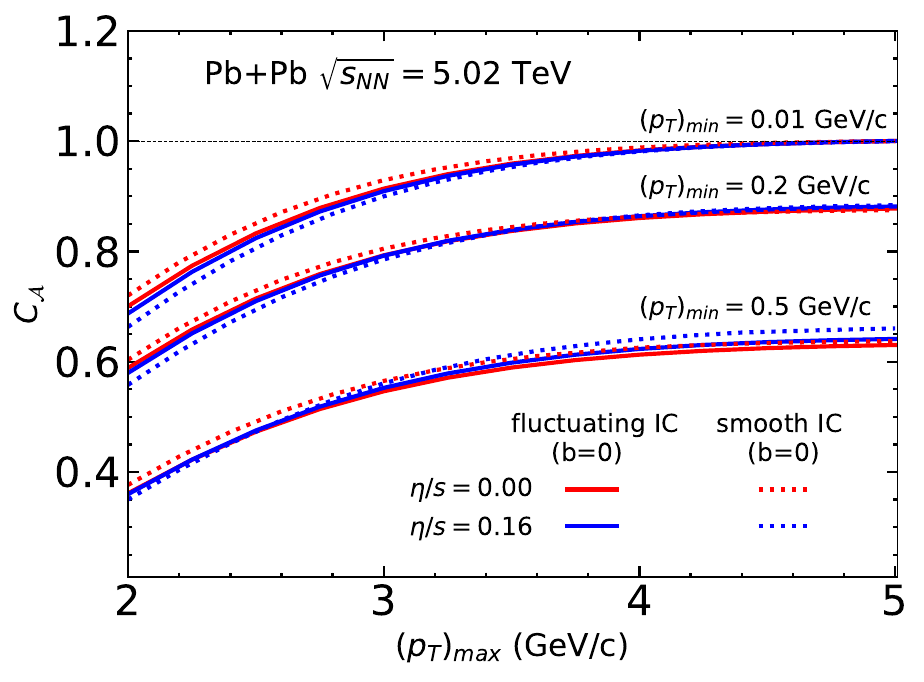} 
	\caption{Variation of the acceptance correction factor (\ref{defCA}) as a function of the upper cut $(p_T)_{\max}$, for various values of the lower cut $(p_T)_{\min}$. } 
	\label{fig:CA}
        \end{figure}
Fig.~\ref{fig:CA} displays the results of our hydrodynamic calculations for $C_{\cal A}$, defined by Eq.~(\ref{defCA}). 
One sees that our predictions are fairly robust, in the sense that they depend little on whether one assumes smooth or fluctuating initial conditions, and on the value of the shear viscosity.
Note that we do not include calculations with bulk viscosity for this quantity. 
The reason is that the numerical results are not reliable for $p_T>3$~GeV$/c$. 
The bulk viscous correction to the equilibrium distribution function in the Cooper-Frye formula causes the $p_T$ spectra to turn negative in the intermediate $p_T$ region~\cite{McNelis:2021acu}. 

Finally, the calculation of $C_{\cal A}$ displayed in Fig.~\ref{fig:CA} is carried out for central collisions. 
But the centrality dependence is negligible at least in the $0-30\%$ centrality range, as all the quantities entering  Eq.~(\ref{defCA}) (scaled spectrum, $\langle p_T\rangle_{\cal A}$, $v_0(p_T)/v_0$), are constant. 

%The dependence of $C_{\cal A}$ on the upper $p_T$ cut is surprisingly large at first sight, since there are very few particles above $p_T>2$~GeV (their number is actually underestimated by the hydrodynamic calculation).
%But these  high-$p_T$ particles get a larger weight in Eq.~(\ref{defCA}).

\section{Applications}
\label{s:applications}

We now illustrate the relevance of the acceptance correction $C_{\cal A}$, and thus of $v_0(p_T)$, in various contexts.  

\subsection{ATLAS data on $\sigma_{p_T}$}
\label{s:ATLAS}

    \begin{figure}[th!]
	\includegraphics[width=\linewidth]{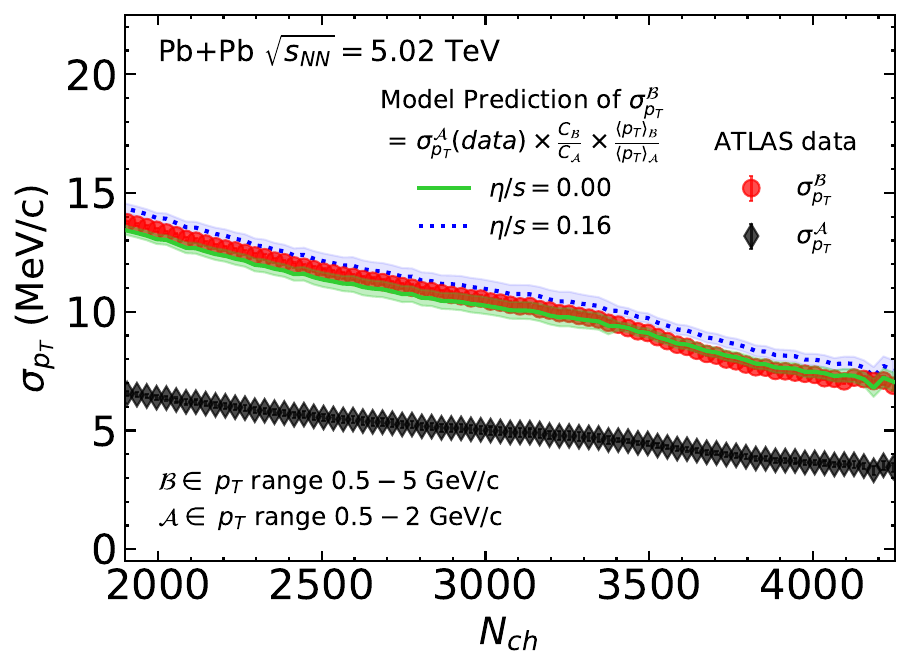} 
	\caption{Symbols: Centrality dependence of $\sigma_{p_T}$ in Pb+Pb collisions at $\sqrt{s_{\rm NN}}=5.02$~TeV measured by ATLAS~\cite{ATLAS:2022dov} in two different transverse momentum windows, $0.5<p_T<2$~GeV$/c$ and $0.5<p_T<5$~GeV$/c$, labeled ${\cal A}$ and ${\cal B}$, respectively.
          Lines: hydrodynamic calculation for $\sigma_{p_T}^{\cal B}$ using as input the experimental result for $\sigma_{p_T}^{\cal A}$ and our hydrodymamic calculations for the other factors ($\langle p_T\rangle$ and $C_{\cal A}$) entering Eq.~(\ref{ptacc2}). 
}
	\label{fig:ATLAS}
    \end{figure}
    The ATLAS collaboration has measured $\sigma_{p_T}$ in several transverse momentum windows~\cite{ATLAS:2022dov}, as a function of the charged particle multiplicity $N_{ch}$. 
    It is surprising at first sight that $\sigma_{p_T}$ increases roughly by a factor two depending on whether or not one includes particles with $p_T>2$~GeV$/c$, as shown in Fig.~\ref{fig:ATLAS}, as these particles constitute only $\sim 2\%$ of the total~\cite{ALICE:2018vuu}. 
Remarkably, this increase is quantitatively reproduced with great precision by our hydrodynamic calculation using Eq.~(\ref{ptacc2}) and the value of $C_{\cal A}$ displayed in Fig.~\ref{fig:CA}. 
Note that the value of $C_{\cal A}$ has been calculated for head-on collisions, but its dependence on centrality (hence on $N_{ch}$) is negligible in the considered range. 

\subsection{PHENIX data on $F_{p_T}$}
\label{s:PHENIX}

    \begin{figure}[th!]
	\includegraphics[width=\linewidth]{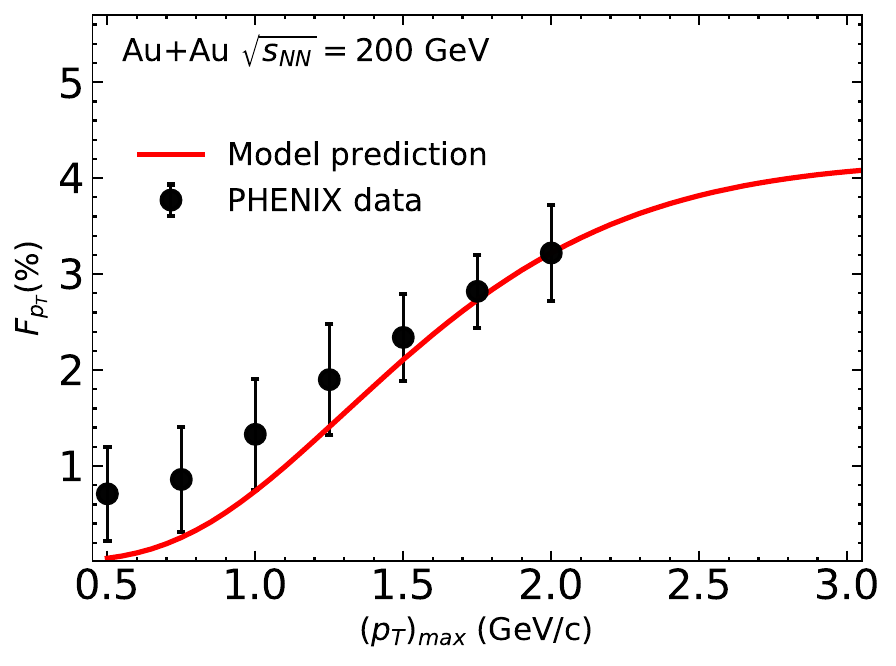} 
	\caption{
 Dependence of $F_{p_T}$ on the upper momentum cut in Au+Au collisions at $\sqrt{s_{\rm NN}}=200$~GeV.
 The lower momentum cut for this analysis is $(p_T)_{\min}=0.2$~GeV$/c$. 
 Symbols: PHENIX data~\cite{PHENIX:2003ccl}.  
 Line: Our hydrodynamic calculation. 
}
	\label{fig:PHENIX}
    \end{figure}
    The PHENIX collaboration has also studied the dependence of momentum fluctuations as a function of the upper $p_T$ cut~\cite{PHENIX:2003ccl}.
They measure the transverse momentum per particle $[p_T]$ in each event: 
\begin{equation}
[p_T]\equiv\frac{1}{N} \sum_{j=1}^N p_{Tj},
\end{equation}
where the sum runs over the $N$ particles seen in the detector, with transverse momenta $p_{Tj}$. 
They then define the relative fluctuation $\omega_{p_T}$ as: 
\begin{equation}
\label{defomegapt}
\omega_{p_T}\equiv\frac{\left(\langle [p_T]^2\rangle-\langle [p_T]\rangle^2\right)^{1/2}}{\langle [p_T]\rangle},
\end{equation}
where angular brackets denote the average over events in a centrality class. 

The quantity $\omega_{p_T}$ is calculated for real events (data), and also for mixed events constructed by randomly mixing tracks from different events, which are uncorrelated. 
Let us first translate this quantity into modern notations. 
Assuming for simplicity that all events have the same multiplicity $N$, simple algebra gives:
\begin{eqnarray}
 \omega_{p_T,{\rm mixed}}&=&\frac{1}{\langle p_T\rangle}\left(
 \frac{\langle p_T^2\rangle-\langle p_T\rangle^2}{N}\right)^{1/2}\cr
  \omega_{p_T,{\rm data}}&=&\frac{1}{\langle p_T\rangle}\left(
 \frac{\langle p_T^2\rangle-\langle p_T\rangle^2}{N}+\sigma_{p_T}^2\right)^{1/2}. 
\end{eqnarray}
Finally, $F_{p_T}$ is defined as the (small) relative difference between $\omega_{p_T,{\rm data}}$ and $\omega_{p_T,{\rm mixed}}$. 
Expanding to leading order in $\sigma_{p_T}^2$, one obtains: 
\begin{eqnarray}
F_{p_T}&\equiv& \frac{\omega_{p_T,{\rm data}}-\omega_{p_T,{\rm mixed}}}{\omega_{p_T,{\rm mixed}}}
\cr
&=&
\left(\frac{\sigma_{p_T}}{\langle p_T\rangle}\right)^2
\frac{N}{2}
\frac{\langle p_T\rangle^2}{\langle p_T^2\rangle-\langle p_T\rangle^2}. 
\end{eqnarray}
All the terms in the right-hand side depend on $p_T$ cuts: 
That of the first term is given by Eq.~(\ref{ptacc2}), while $N$, $\langle p_T\rangle$ and $\langle p_T^2\rangle$ must be evaluated by integrating the average spectrum $N_0(p_T)$ with the proper bounds. 

We now compare these data with hydrodynamic calculations.
While evaluating $\sigma_{p_T}$ requires a sophisticated model of initial conditions, its dependence on the momentum cuts is simply given by Eq.~(\ref{ptacc2}).
In order to evaluate this dependence, we repeat the smooth hydrodynamic calculation of $v_0(p_T)/v_0$, rescaling the initial entropy density by $s_{\rm NN}^{0.155}$~\cite{ALICE:2015juo} in order to take into account the different collision energy~\cite{Gardim:2024zvi}.\footnote{We implement a shear viscosity over entropy ratio $\eta/s=0.08$, but the result displayed in Fig.~\ref{fig:PHENIX} depends very little on $\eta/s$.} 
Our hydrodynamic calculation is still for central collisions, while the data displayed in Fig.~\ref{fig:PHENIX} are in the $20-25\%$ centrality window.
We again make use of the property that $C_{\cal A}$ is independent of centrality. 

The result for $F_{p_T}$ is displayed in Fig.~\ref{fig:PHENIX}, where we have normalized our calculation so as to match the rightmost data point.  
Data are above our calculation for low values of $(p_T)_{\max}$, where we expect that $F_{p_T}$ should  essentially vanish. 
A similar discrepancy was observed between these data and event-by-event hydrodynamic calculations~\cite{Bozek:2012fw}. 

The reason is probably that the data are biased by short-range correlations, referred to as ``non-flow'' in the context of anisotropic flow analyses~\cite{Ollitrault:2023wjk}. 
The detector used for this analysis covers a small pseudorapidity window $|\eta|<0.35$, and half of the azimuthal angle. 
At low $p_T$, there are sizable Bose-Einstein correlations between identical pions, referred to as HBT correlations. 
Specifically, for a pion with a given $p_{Tj}$, there is a larger probability of finding another pion with almost the same momentum than in mixed events. 
Nonflow effects induced by HBT correlations have been studied in the context of anisotropic flow~\cite{Dinh:1999mn} and are largely suppressed by implementing a small pseudorapidity gap $|\Delta\eta|>0.1$~\cite{STAR:2000ekf}. 
This example illustrates that the analysis of $[p_T]$ fluctuations should be carried out by isolating a long-range correlation, as explained in Appendix~\ref{s:analysis}. 

\subsection{Determination of the speed of sound}
\label{s:cs}

Finally, the acceptance factor $C_{\cal A}$ is also relevant when analyzing the increase of the mean transverse momentum in ultracentral collisions, from which one infers the speed of sound in QGP~\cite{CMS:2024sgx}.
The physics context is different: 
Instead of studying the event-by-event fluctuations of $[p_T]$ at fixed multiplicity, one studies how the average of $[p_T]$ over events increases as a function of the event multiplicity.
But Eq.~(\ref{deltanpt2}), on which our derivation is based, applies to both cases. 
Therefore, Eq.~(\ref{ptacc}) shows that the relative increase of $\langle p_T\rangle$ is multiplied by a factor $C_{\cal A}$ when $p_T$ cuts are implemented. 
In the equation used to determine the speed of sound: 
\begin{equation}
\label{cs2}
c_s^2\equiv \frac{d\ln \langle p_T\rangle}{d\ln N_{ch}},
\end{equation}
the numerator gets a correction factor $C_{\cal A}$ due to momentum cuts: 
\begin{equation}
d\ln\langle p_T\rangle_{\cal A}=C_{\cal A} \, d\ln\langle p_T\rangle.
\end{equation}
If no extrapolation of $p_T$ spectra is carried out before applying Eq.~(\ref{cs2}), then the estimate of $c_s^2$ should be divided by $C_{\cal A}$. In Ref.~\cite{CMS:2024sgx}, CMS uses an extrapolation in order to access the $p_T$-spectra at the low $p_T$ region and then they use it for the extraction of $c_s^2$. The usefulness of the correction factor $C_A$ is that CMS can measure $c_s^2$ in their accessible $p_T$-acceptance range and then divide by the corresponding $C_A$ for the entire $p_T$-range to obtain the correct $c_s^2$. This would serve as a nice check for the validity of the correction factor $C_A$ and would also remove uncertainties due to extrapolation.
Fig.~\ref{fig:CA} shows that even modest lower and upper cuts $0.2<p_T<3$~GeV$/c$ imply a correction of the order of 25\%. 

\section{Conclusions}
\label{s:summary}

$v_0(p_T)$ is a new observable which can be used to  probe collective behavior in ultrarelativistic nucleus-nucleus collisions, and which is just as important as anisotropic flow, $v_n(p_T)$. 
It also comes from a long-range correlation, and displays the same characteristic $p_T$ dependence and mass ordering, which originate from the boosted thermal distribution. 
One often speaks of ``radial flow'' when fitting $p_T$-spectra of identical particles with blast-wave models~\cite{Schnedermann:1993ws}. 
But in reality, $v_0(p_T)$ is the {\it true radial flow\/}, in the sense that it would vanish if there was no collective motion, in the same way as elliptic flow. 

The ``integrated'' flow $v_0$, which is equivalent to the well-known observable $\sigma_{p_T}$, carries most of the information about the initial state, and also most of the sensitivity to transport coefficients, like the integrated anisotropic flow $v_n$. 
We have chosen to study the scaled flow $v_0(p_T)/v_0$, which singles out the differential properties, and which we predict to have little dependence on system size and centrality at a given collision energy, like $v_n(p_T)/v_n$. 
We have made quantitative hydrodynamic predictions for $v_0(p_T)/v_0$ at LHC energies. 
%We have observed that it depends significantly on the bulk viscosity, but little on the shear viscosity, while anisotropic flow is equally sensitive to both~\cite{Gardim:2022vys}. 
%A combined analysis of $v_0(p_T)$ and $v_n(p_T)$ should help disentangling the effects of shear and bulk viscosities. 

We have shown that $v_0(p_T)$ is also the quantity which determines the dependence on $p_T$ cuts of the increase of $\langle p_T\rangle$ in ultracentral collisions, and of $\sigma_{p_T}$. 
Our hydrodynamic calculations are in quantitative agreement with recent ATLAS data, and in reasonable agreement with older PHENIX data. 
We anticipate that detailed measurements of $v_0(p_T)$ will soon be available at the LHC, and will provide new constraints for hydrodynamic models. 

A next step will be to study the rapidity decorrelation~\cite{CMS:2015xmx,ATLAS:2017rij,ALICE:2023tvh}, which should affect $v_0(p_T)$ in a similar way as $v_n(p_T)$. 
On the theoretical side, this requires to model the breaking of boost invariance in hydrodynamic calculations~\cite{Shen:2017bsr,Bozek:2017qir,Pang:2018zzo,Zhu:2024tns}. 
Finally, in the same way as anisotropic flow has been observed in proton-nucleus~\cite{Bozek:2011if,CMS:2012qk,ALICE:2012eyl,ATLAS:2012cix,CMS:2019wiy} and proton-proton~\cite{CMS:2010ifv,ATLAS:2015hzw,CMS:2016fnw} collisions, one expects $v_0(p_T)$ to be also present in small systems, where it will provide a new probe of collectivity~\cite{Nagle:2018nvi}. 

\begin{acknowledgments}
JYO thanks Giuliano Giacalone, Bedangadas Mohanty, Bjoern Schenke, Chun Shen, Swati Saha and Derek Teaney for discussions. RS thanks Piotr Bo\.zek for discussions on shear and bulk viscosity. RS is supported by the Polish National Science Centre grant: 2019/35/O/ST2/00357.
\end{acknowledgments}

\appendix
\section{Experimental procedure}
\label{s:analysis}

First, one should carry out the usual centrality selection. 
Note that $v_0(p_T)$ was originally defined by considering a set of events with the exact same multiplicity~\cite{Schenke:2020uqq}. 
However, the centrality selection is often done using a separate detector, so that the multiplicity fluctuates at fixed centrality. 
But these fluctuations are unlikely to affect the determination of $v_0(p_T)$ for the following reason: 
The scaled $p_T$ spectrum and $\langle p_T\rangle$ are almost constant up to $30\%$ centrality, so that their ($\delta N(p_T)$ and $\delta p_T$) event-by-event fluctuations (which $v_0(p_T)$ quantifies) within a centrality bin, are not generated by variations of the impact parameter. 
Therefore, it is fine to work with a range of event multiplicities, and it is probably not crucial to use a very narrow centrality window. 

The analysis of $v_0(p_T)$ should be similar to that of anisotropic flow $v_n(p_T)$.
That is, it should be inferred from a long-range correlation~\cite{PHENIX:2003qra}, so as to suppress nonflow effects~\cite{Zhang:2021phk}. 
This implies that $\delta N(p_T)$ and $\delta p_T$ in Eq.~(\ref{defv0pt}) should be measured in different pseudorapidity windows. 
Similarly, the fluctuation $\sigma_{p_T}$ should be measured by correlating the transverse momenta of particles~\cite{STAR:2005vxr,ALICE:2014gvd,ATLAS:2019pvn} in different windows. 

The simplest is to use two ``subevents'', corresponding to two different pseudorapidity windows ${\cal A}$ and ${\cal B}$, separated by a gap $\Delta\eta$. 
We denote by ${\cal A}$ the window where $v_0(p_T)$ is analyzed, and by ${\cal B}$ the other window, used as a reference~\cite{Luzum:2012da}. 
For each event, one calculates the following quantities:
\begin{itemize}
\item{The fraction $f(p_T)$ of particles in ${\cal A}$ for each $p_T$ bin.\footnote{We recommend to use the event-by-event fraction of particles in the bin, rather than just the bin multiplicity, so as to avoid effects of multiplicity fluctuations, but both are probably equivalent.}
}
\item{The transverse momentum per particle in ${\cal A}$ and ${\cal B}$, denoted by $[p_T]_{\cal A}$ and $[p_T]_{\cal B}$.} 
\end{itemize}
Then, $\sigma_{p_T}$ is defined as:
\begin{equation}
  \label{sigmaptexp}
  \sigma_{p_T}^2\equiv \langle [p_T]_{\cal A}[p_T]_{\cal B}\rangle-\langle [p_T]_{\cal A}\rangle\langle [p_T]_{\cal B}\rangle, 
\end{equation}
where angular brackets denote an average over events in the centrality class. 
This implicitly assumes that $\sigma_{p_T}$ is the same for ${\cal A}$ and ${\cal B}$, i.e., that subevents are equivalent (which is strictly the case for subevents located symmetrically around $\eta=0$), and that the rapidity decorrelation~\cite{CMS:2015xmx,ATLAS:2017rij,ALICE:2023tvh} is negligible.  
The integrated flow $v_0$ is then defined using Eq.~(\ref{defv0}). 
In which detector $\langle p_T\rangle$ is measured is a matter of convention. 
Finally, $v_0(p_T)$ is then obtained through:
\begin{equation}
  v_0(p_T) = \frac{\langle f(p_T)[p_T]_{\cal B}\rangle-\langle f(p_T)\rangle\langle[p_T]_{\cal B}\rangle}{\langle f(p_T)\rangle\sigma_{p_T}}.
  \end{equation}
It satisfies  the sum rules: 
\begin{eqnarray}
  \int_{p_T} v_0(p_T) \langle f(p_T)\rangle 
  &=& 0 \cr
  \int_{p_T}p_T v_0(p_T) \langle f(p_T)\rangle 
 & =& \sigma_{p_T}\int_{p_T} \langle f(p_T)\rangle, 
\end{eqnarray}
where $\int_{p_T}$ denotes the sum over all bins.

\end{document}